# Xabclib : A Fully Auto-tuned Sparse Iterative Solver[i]

**Takahiro Katagiri**
Information Research Center,
The University of Tokyo
2-11-16 Yayoi, Bunkyo-ku,
Tokyo 113-8658, JAPAN
+81-3-5841-2743

katagiri@cc.u-tokyo.ac.jp

**Takao Sakurai**
Central Research Laboratory,
Hitachi Ltd.
292 Yoshida-cho, Totsuka-ku,
Yokohama, Kanagawa 244-0817,
JAPAN
+81-45-860-2142

takao.sakurai.ju@hitachi.com

**Mitsuyoshi Igai**
Hitachi ULSI Systems Co., Ltd.
292 Yoshida-cho, Totsuka-ku,
Yokohama, Kanagawa 244-0817,
JAPAN
+81-45-860-2144

mitsuyoshi.igai.bf@hitachi.com

**Shoji Itoh**
Information Technology Center,
The University of Tokyo
2-11-16 Yayoi, Bunkyo-ku,
Tokyo 113-8658, JAPAN
+81-3-5841-0794

itosho@cc.u-tokyo.ac.jp

**Satoshi Ohshima**
Information Technology Center,
The University of Tokyo
2-11-16 Yayoi, Bunkyo-ku,
Tokyo 113-8658, JAPAN
+81-3-5841-2737

ohshima@cc.u-tokyo.ac.jp

**Hisayasu Kuroda**
Graduate School of Science and
Engineering, Ehime University
3 Bunkyo-cho, Matsuyama
Ehime 790-8577, JAPAN
+81-89-927-8406

kuroda@cs.ehime-u.ac.jp

**Ken Naono**
Central Research Laboratory,
Hitachi Ltd.
292 Yoshida-cho, Totsuka-ku,
Yokohama, Kanagawa 244-0817,
JAPAN
+81-45-860-2140

ken.naono.aw@hitachi.com

**Kengo Nakajima**
Information Technology Center,
The University of Tokyo
2-11-16 Yayoi, Bunkyo-ku,
Tokyo 113-8658, JAPAN
+81-3-5841-2719

nakajima@cc.u-tokyo.ac.jp

## ABSTRACT

In this paper, we propose a general application programming interface named OpenATLib for auto-tuning (AT). OpenATLib is designed to establish the reusability of AT functions. By using OpenATLib, we develop a fully auto-tuned sparse iterative solver named Xabclib. Xabclib has several novel run-time AT functions. First, the following new implementations of sparse matrix-vector multiplication (SpMV) for thread processing are implemented: (1) non-zero elements; (2) omission of zero-elements computation for vector reduction; (3) branchless segmented scan (BSS). According to the performance evaluation and the comparison with conventional implementations, the following results are obtained: (1) 14× speedup for non-zero elements and zero-elements computation omission for symmetric SpMV; (2) 4.62× speedup by using BSS. We also develop a "numerical computation policy" that can optimize memory space and computational accuracy. Using the policy, we obtain the following: (1) an averaged 1/45 memory space reduction; (2) avoidance of the "fault convergence" situation, which is a problem of conventional solvers.

## Categories and Subject Descriptors

D.2.13 [**Reusable Software**]: Reusable libraries; G.1.3 [**Numerical Linear Algebra**]: Eigenvalues and eigenvectors (direct and iterative methods), Linear systems (direct and iterative methods)

## General Terms

Algorithms, Management, Measurement, Performance, Design, Standardization

## Keywords

OpenATLib, Xabclib, Sparse Matrix-Vector Multiplication (SpMV), Auto-tuning, Branchless Segmented Scan, Numerical Computation Policy, GMRES, BiCGStab, Arnoldi, Lanczos

Current computer architectures are generally too complicated to tune the performance of numerical computations. In addition, the increased number of cores in multi-core architectures, the deep hierarchical caches, and the non-uniform memory accesses deteriorate the performance of crucial processes for numerical computations. For these reasons, the cost of developing numerical software is becoming higher and higher. This situation is now creating a software crisis in high-performance numerical software.

To handle this critical situation, auto-tuning (AT) for numerical processing has been studied at several levels of numerical software. At the lower level, which is the basic numerical linear algebra (BLAS) level, ATLAS [1] is well-known AT software for dense matrix libraries. For sparse matrix libraries, SPARSITY [2] and OSKI [3] offer high performance for sparse matrix-vector multiplication (SpMV). At the higher level, which is numerical algorithms or solvers, FFTW [4] consists of fast Fourier transform routines that are widely used in AT libraries. For sparse iterative libraries, ILIB [5] has interesting AT functions, including MPI optimization, that perform at run-time. For dense matrix libraries, ABCLib [6] provides an AT at the medium level of numerical computation, such as orthogonalization or non-BLAS based computations. In addition, ABCLib also provides new AT timing, called before execution for their AT framework of FIBER.

In this paper, we propose novel AT functions for a fully tuned sparse iterative solver. The novel features from the numerical library point of view are summarized in the next section.

## 1.1 Originalities of This Paper
In this paper, we propose the following new AT functions for sparse iterative libraries. All functions are implemented as a library to provide application programming interfaces (APIs) for dedicated AT functions.

First, we propose a general API named OpenATLib for the AT framework. OpenATLib is designed to establish the reusability of AT functions. This is the first originality of this paper.

Second, we develop new implementations of high-performance sparse matrix-vector multiplication (SpMV) with OpenMP. We propose the following new thread implementations: (1) non-zero elements based decomposition; (2) omission of zero-elements computation for vector reduction; (3) branchless segmented scan (BSS) for scalar multi-core architectures.

Third, we develop a "numerical computation policy" for sparse iterative solvers. The conventional ATs focus on execution time optimization only. We propose an AT strategy to optimize memory space and accuracy as functions of the numerical computation policy, based on the end-user requirements.

Fourth, we implement a general AT strategy to adjust the parameters of numerical algorithms at run-time. The strategy is based on numerical algorithm independence, which means the strategy is based only on monitoring crucial values in iterative solvers. The strategy does not need linear algebra theory, and hence it can be widely used for numerical algorithms.

## 1.2 Organization
This paper is organized as follows. In Section 2, OpenATLib and Xabclib are explained. Section 3 explains the AT strategy used in OpenATLib. Section 4 is a performance evaluation with one node of the T2K Open Supercomputer (Todai combined cluster), which consists of 16 cores with 4 sockets of the AMD Opteron. Section 5 gives the conclusion of this paper.

## 2. OpenATLib AND Xabclib
## 2.1 Design Space for OpenATLib
OpenATLib is a general API library for the AT functions. OpenATLib focuses on AT functions for constructing a sparse iterative solver using the Krylov subspace method. Based on OpenATLib, Xabclib is the reference implementation of the auto-tuned sparse numerical library. In this section, we explain the design space of OpenATLib.

OpenATLib is designed to supply APIs to select the best run-time implementation for numerical library developers. The design space provides the following.

- To provide the algorithm selection:
  - Restart frequency for the Krylov subspace method.
  - Re-orthogonalization algorithms.
- To provide the implementation selection:
  - Sparse matrix-vector multiplication (SpMV)

- Branchless segmented scan (BSS).
- Load-balanced row-based decomposition method (non-zero elements based method)
- Zero-elements computation omission for vector reduction.
- To provide the computer resource selection:
  - SpMV with respect to the memory space and the number of cores.
- To provide the numerical computation policy from end-users:
  - Execution speed optimization.
  - Memory space optimization.
  - Computational accuracy optimization.

## 2.2 Details of Supplied Functions
Due to the Krylov subspace method, OpenATLib has enough functions to optimize implementations of SpMV. The specific parameters for numerical algorithms can also be optimized.

The functions provided by OpenATLib are classified as the following four categories:

1. Computation function (e.g., OpenATI_DSRMV)
2. Auxiliary function (e.g., OpenATI_DAFRT)
3. Setup function (e.g., OpenATI_INIT, OpenATI_DSRMV_Setup)
4. Meta-library interface (e.g., OpenATI_LINEARSOLVE)

The names of these functions begin with "OpenATI_" and end with a descriptive term. The function names and their descriptions are summarized in Table 1.

**Table 1. Functions Details for OpenATLib.**

| Function Name | Description |
|---|---|
| OpenATI_INIT | Set default parameter for OpenATLib and Xabclib. |
| OpenATI_DAFRT | Judge increment for restart frequency on the Krylov subspace. |
| OpenATI_DSRMV | Judge the best implementation for double precision symmetric SpMV on CRS format. |
| OpenATI_DURMV | Judge the best implementation for double precision non-symmetric SpMxV on CRS format. |
| OpenATI_DSRMV_Setup | Setup function for OpenATI_DSRMV. |
| OpenATI_DURMV_Setup | Setup function for OpenATI_DURMV. |
| OpenATI_DAFGS | Gram-Schmidt re-orthogonalization functions with 4 implementations. |
| OpenATI_LINEARSOLVE | Numerical computation policy interface for linear equations |

| | solver. |
|---|---|
| OpenATI_EIGENSOLVE | Numerical computation policy interface for eigensolver. |

Table 1 lists two interfaces, OpenATI_LINEARSOLVE and OpenATI_EIGENSOLVE, for the numerical computation policy. Because these two interfaces include several APIs for OpenATLib at lower levels, we refer to each as a "meta-interface."

In the current implementation, we do not implement a solver selection function for the meta-solvers. Instead, we implement the following solver candidates.

- OpenATI_LINEARSOLVE
  - Lanczos: a restarted Lanczos [7].
  - Arnoldi: a restarted Arnoldi [8].
- OpenATI_EIGENSOLVE
  - GMRES: a restarted GMRES(m) [9].
  - BiCGStab: a preconditioned BiCGStab modified by Itoh [10].

## 2.3 Numerical Computation Policy

### 2.3.1 Overview
One of the original functions of OpenATLib is the numerical computation policy. This function determines the AT policy from the end-user requirements [14]. The current implementation establishes AT functions based on execution time, memory space, and computational accuracy.

To use the numerical computation policy function, OpenATLib supplies the following APIs. OpenATI_LINEARSOLVE is designed for a linear equations solver with unsymmetric real matrices. OpenATI_EIGENSOLVE is designed for a standard eigenproblem solver with symmetric and unsymmetric real matrices.

OpenATI_LINEARSOLVE and OpenATI_EIGENSOLVE use the set optimized arguments of the Xabclib implementation, based on the end-user's numerical policy.

### 2.3.2 Usage of the Numerical Computation Policy
When end-users use the meta-solvers, they make a "numerical policy file" in the current directory. The name of the file is "OPENATI_POLICY_INPUT.#", where # indicates the identifier of the threads to be set by the policy.

The format of the policy file is as follows.

```
<keyword> = <value>
```

The words POLICY, CPU, RESIDUAL, MAXMEMORY, MAXTIME, PRECONDITIONER, SOLVER are configurable keywords. An unregistered <keyword> in the policy file is the default value "TIME". The <keyword> is as follows.

```
POLICY = <value>

<value>: TIME / ACCURACY / MEMORY / STABLE
```

"TIME" is the default policy. The details of the policy definition are summarized as follows.

1. If POLICY=TIME, the meta-solvers prioritize execution time over accuracy and saving memory. Implementation for higher performance is selected.

2. If POLICY=MEMORY, the meta-solvers set arguments to establish the least possible memory usage.

3. If POLICY=ACCURACY, the meta-solvers recalculate the errors of the solution vector after it has converged. If the computed error does not satisfy the end-user requirement, the meta-solvers call the meta-solver again. Then, the policy performs the computation until the end-user requirement is satisfied or the allowed maximum execution time has passed.

4. If POLICY=STABLE, the meta-solvers set arguments without the AT. This is used for debugging end-user software.

## 3. AUTO-TUNING STRATEGY

### 3.1 Overview
The AT implemented in OpenATLib is fully run-time. This is because it detects numerical information for input sparse matrices, such as the distribution of non-zero elements and the numerical properties matching the numerical algorithm used.

To establish the fully run-time AT, we need two kinds of AT timings: (1) before execution; (2) in the inner iterative loop. For (1), the AT performs before the iterations for the sparse iterative solver. For example, measuring SpMV time is the primary process. For (2), the AT performs in every iteration. For example, adjustment of the re-start frequency is a typical process of (2). In OpenATLib, the candidates of SpMV implementations have a maximum of 4 iterations whereas the number of iterations for a numerical algorithm is 100 or more. The overhead of (1), here, is negligible compared to the main iteration part of the sparse solver.

The inner iteration loop in (2) is classified into two kinds of iterations: (a) inner iteration; (b) outer iteration. This is due to the numerical computation policy, especially the accuracy policy. Inner iteration (a) is performed by the numerical algorithm. The advantage here is that the error computation for the solution vector in the inner iteration is not implemented in the loop of (a), since this computation causes very high overhead. In many numerical solvers, residual information from the theory is kept. However, the residual information is not always equal to the "real" residual information. In this case, the accuracy requirement from the end-user may not be satisfied. We call this situation "fault convergence." To solve the fault convergence situation, we use the outer loop iteration (b). This establishes the accuracy policy. With the accuracy policy, the required accuracy from the end-user is retained if the matrix converges. However, the execution time increases compared to that of the other policies. The whole process of the AT is shown in Figure 1.

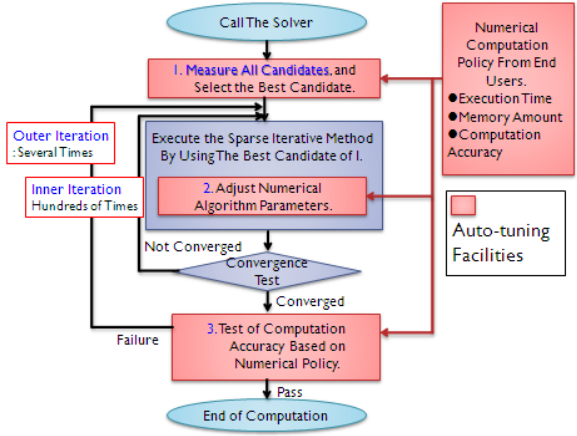

**Figure 1. The Process Flow of Run-time Auto-tuning.**

## 3.2 Krylov Subspace Parameter Optimization

Examples of the Krylov subspace method include the Lanczos method for eigensolvers and the GMRES method for linear equation solvers. These methods must specify the available maximum dimension for the inner Krylov subspace. In many cases, the end-users fix the dimension according to the available maximum memory space.

If the number of iterations is more than the fixed dimension, a new computation, beginning with the current calculated approximation as an initial vector, is performed in order to span the new Krylov subspace. This process is called a "restart." The number of iterations is called the "restart frequency." If the restart frequency is too small, stagnation occurs and causes reduction of the residual vector. To eliminate stagnation, the total number of iterations is increased. In most cases, this computation is for re-orthogonalization, and, as a result, the execution time is very much increased.

The best frequency depends on the numerical condition of the input sparse matrix. However, it is very difficult to estimate the best frequency without execution. In other words, from the library point of view, we need to perform the run-time AT to set the frequency.

OpenATI_DAFRT enables library developers to judge the restart frequency based on current information of the Krylov subspace.

### 3.2.1 Strategy of Restart Frequency Auto-tuning

The method shown in this paper operates independently of the numerical algorithms, since the method only monitors the residual vector at run-time. The principal idea is as follows.

The norm of stagnation is defined as the maximum value divided by the minimal value from the $t$-th time to the $s$-th time. This value is called the "ratio of the max-min in the residual." For simplification, we call this the "MM ratio."

The MM ratio to the past $t$-th time, namely $R_i$ $(s,t)$, can be described with the $i$-th residual $r_i$, as shown in the following formula.

$$R_i(s,t) = \frac{\max_z \{r_i(z); z = s-t+1, \cdots, s\}}{\min_z \{r_i(z); z = s-t+1, \cdots, s\}} \quad \dots(1)$$

If the restart frequency is large enough, the residual tends to greatly reduce in value; hence, the MM ratio is large. Conversely, if the restart frequency is small, it tends to cause stagnation; hence, the MM ratio is small. We can control the restart frequency at run-time by monitoring the sequence of values in equation (1).

**Example: Using the API for the Restart Frequency**

We can use the function OpenATI_DAFRT in arbitrary programs. In this section, an example of the implementation is shown in Figure 2.

In the example of Figure 2, the judgment of the restart frequency is 5 iterations. If this number needs to increase, the frequency is increased by 1. If the library developers want to write such an AT for their programs, they can write code similar to that shown in Figure 2. This strategy is implemented in Xabclib.

### 3.2.2 SpMV Implementation Selection

It is a well-known fact that the heaviest process in many algorithms based on the Krylov subspace is SpMV. Hence, we supply multiple candidates to perform AT for SpMV in OpenATLib. The SpMV is parallelized with OpenMP.

Our originalities for SpMV implementations are based on the following three new implementations.

First, we carry out load-balanced implementation based on the number of non-zero elements per row. This implementation has merit for handling sparse matrices that include a very imbalanced number of non-zero elements per row. The conventional implementation only supports parallelism for each row and does not consider the load balancing caused by the number of non-zero elements per row. Figure 3 shows the case of unbalancing.

```
// Parameter Definition

MSIZE=1  //Initial restart frequency.

I=5      //Judgment frequency.

        --- Omission ---

IF RSDID < TOL  RETURN  //Convergence Test

SAMP (K)=RSDID  //Set residual to SAMP(K).

IF (mod (K, I) .eq. 0) THEN  //Call DAFRT per I times.

   IRT=0

   CALL  OpenATI_DAFRT (I, SAMP, IRT,
                IATRARAM, RATPARAM, INFO)

   IF IRT= 1  MSIZE=MSIZE+1

        //Increase restart frequency.

   K=0

END IF

K=K+1

        --- Omission ---
```

**Figure 2. Example of OpenATI_DAFRT.**

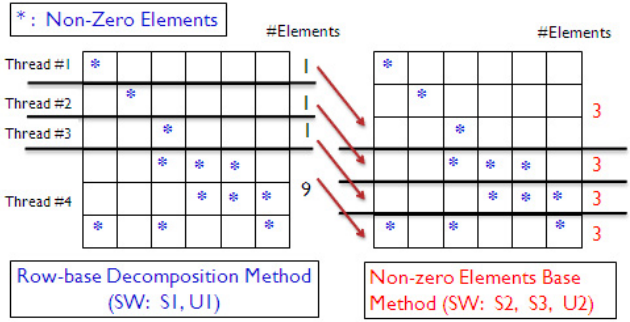

**Figure 3. Example matrix causing heavy unbalanced computations due to the number of non-zero elements. This also shows the effect of parallelization for non-zero elements.**

Second, we modify the original segmented scan (SS) method [11] for SpMV to match scalar architectures. The method is named a branchless segmented scan (BSS) [12]. The conventional SS method was developed for vector architectures [11], and so it performs poorly in scalar architectures. We removed the IF-statements located in the innermost loop in the original SS. By introducing a new data format representing the continuous location of the source arrays of SpMV, we can remove the IF-statements [12].

Third, for symmetric SpMV, we reduce the redundant operations in the threads. If we handle almost tridiagonal structures or matrices with stencil computation, most of the reductions are performed on zero values. These do not contribute to actual FLOPS computations. To eliminate this situation, we search non-zero regions for the vector of reduction in advance, and then the regions are stored in a special data structure. This method is very crucial in such matrices, since the conventional implementations cannot be scaled according to the number of threads, unlike our method, which can be scaled in several matrices.

In summary, OpenATLib is implemented for the following AT candidates of SpMV:

- OpenATI_DSRMV (A symmetric SpMV)
  - S1) Row-based decomposition method (RDM).
  - S2) Non-zero elements based method (Load-balancing RDM).
  - S3) Non-zero elements based method with zero-element computation omission for vector reduction.
- OpenATI_DURMV (An unsymmetric SpMV)
  - U1) RDM.
  - U2) Load-balancing RDM (for scalar multi-core processors).
  - U3) Branchless segmented scan (BSS) (for scalar multi-core processors).
  - U4) Original segmented scan (for vector processors).

In addition to the AT of SpMV, OpenATLib supplies basic operations for numerical linear algebra. Re-orthogonalization is one of the important processes in numerical linear algebra. The

AT for re-orthogonalization performs according to the accuracy policy in the current implementation, and the AT of the inner iteration level for re-orthogonalization can be adapted. A focus of future work is to extend the current functions in OpenATLib. OpenATLib supports the following four kinds of implementations.

- To provide re-orthogonalization:
  - Classical Gram-Schmidt method (CGS)
    - ◇ Accuracy of re-orthogonalization is lower. Parallel performance is excellent.
  - DGKS method [13, 14]
    - ◇ Supplies improved accuracy by running CGS two times.
  - Modified Gram-Schmidt method (MGS): The default algorithm
    - ◇ Most effective performance and accuracy in many cases.
  - Blocked classical Gram-Schmidt method (BCGS)
    - ◇ Re-orthogonalization by an intra-block with CGS, and by an inter-block with MGS. The block length is 4.

## 3.3 Numerical Computation Policy Function

### 3.3.1 Overview
A conventional AT focuses on execution speed only. However, some applications require small memory space or high computational accuracy over execution time. To extend the AT to such applications, we implemented the "numerical computation policy." This function is activated by the end-user.

The strategy for each policy is important to construct practical AT functions. Our run-time AT is based on measuring all candidates for the implementations; hence, it is easy to make a strategy for the execution time policy. The memory space policy is also easy because we know the memory amounts for all OpenATLib implementations.

The accuracy policy is very difficult to implement; since many approaches are possible in the construction of the policy. Imagine that we can use multiple precision arithmetic routines or iterative refinement algorithms. In our strategy, we do not use a multiple precision arithmetic or iterative refinement algorithm. Instead, we focus on the "fault convergence" situation, as in the case of using the Krylov subspace method.

### 3.3.2 Accuracy Policy Strategy
To solve the fault convergence situation, we separate the iterative loop into two loops: outer and inner. In the inner loop, we use only residual vector information without direct computation. After convergence, we compute the "real" residual error by using a solution vector. If the "real" residual error does not satisfy the end-user requirements, then the computation is performed again with a narrowed threshold for the convergence test. This is the basic strategy behind our accuracy policy.

The re-orthogonalization algorithm affects the convergence in general. In addition, the algorithm also affects the parallel performance. As the trade-off between accuracy and parallel execution speed for re-orthogonalization, we define the following strategy.

- If the end-user accuracy requirement is less than or equal to $10^{-10}$, use MGS. Otherwise, use BCGS.

- If fault convergence occurs, use DGKS.

This strategy is based on empirical evidence that the DGKS behavior is more stable than that of the MGS in some situations. Figure 4 shows the process flow of this accuracy policy.

# 4. PERFORMANCE EVALUATION

## 4.1 Machine Environments

We used the T2K Open Supercomputer (TODAI Combined Cluster), known by the production name HITACHI HA8000, installed in the Information Technology Center at the University of Tokyo. Each node contains four sockets of the AMD Opteron 8356 (Quad core, 2.3 GHz). The L1 cache is 64 KB/core, the L2 cache is 512 KB/core, and the L3 cache is 2 MB/4 cores. The memory on each node is 32 GB with the DDR2-667 MHz. The theoretical peak is 147.2 GFLOPS/node. We used Intel Fortran Compiler Professional Version 11.0 with the option "-O3 -m64 -openmp -mcmodel=medium."

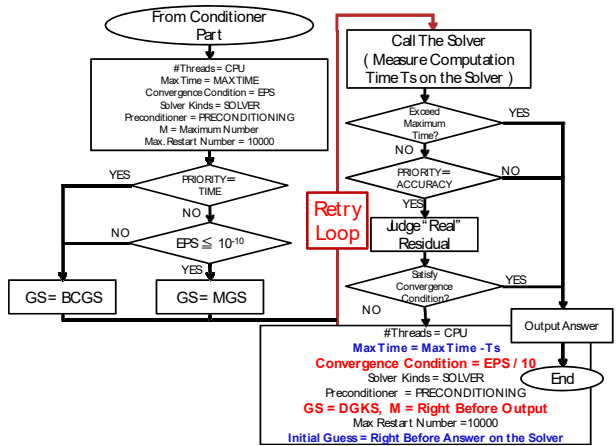

**Figure 4. Process Flow of the Accuracy Policy.**

## 4.2 Test Matrices

We used 21 kinds of symmetric matrices and 22 kinds of unsymmetric matrices from the University of Florida sparse matrix collection (hereafter, "UF collection") [15]. Information of the UF collection, including the matrix name, the dimension, and the number of non-zero elements is shown in Table 2 and Table 3.

**Table 2. Test Matrices (Symmetric).**

| No. | Matrix Name | Dimension | #Non-zeros |
|-----|-------------|-----------|------------|
| 1 | vibrobox | 123,328 | 177,578 |
| 2 | Lin | 256,000 | 1,011,200 |
| 3 | cfd1 | 70,656 | 949,510 |
| 4 | cfd2 | 123,440 | 1,605,669 |
| 5 | gyro | 17,361 | 519,260 |
| 6 | t3dl | 20,360 | 265,113 |
| 7 | c-71 | 76,638 | 468,096 |
| 8 | c-73 | 169,422 | 724,348 |
| 9 | Si5H12 | 19,896 | 379,247 |
| 10 | SiO | 33,401 | 675,528 |
| 11 | dawson5 | 51,537 | 531,157 |
| 12 | H2O | 67,024 | 1,141,880 |
| 13 | F2 | 71,505 | 2,682,895 |
| 14 | oilpan | 73,752 | 1,835,470 |
| 15 | shipsec1 | 140,874 | 3,977,139 |
| 16 | bmw7st_1 | 141,347 | 3,740,507 |
| 17 | SiO2 | 155,331 | 5,719,417 |
| 18 | shipsec5 | 179,860 | 5,146,478 |
| 19 | Si4Ge41H72 | 185,639 | 7,589,452 |
| 20 | bmw3_2 | 227,352 | 5,757,996 |
| 21 | Ga41As41H72 | 268,096 | 9,378,286 |

**Table 3. Test Matrices (Unsymmetric)**

| No. | Matrix Name | Dimension | #Non-zeros |
|-----|-------------|-----------|------------|
| 1 | chipcool0 | 20,082 | 281,150 |
| 2 | chem_master1 | 40,401 | 201,201 |
| 3 | torso1 | 116,158 | 8,516,500 |
| 4 | torso2 | 115,067 | 1,033,473 |
| 5 | torso3 | 259,156 | 4,429,042 |
| 6 | memplus | 17,758 | 126,150 |
| 7 | ex19 | 12,005 | 259,879 |
| 8 | poisson3Da | 13,514 | 352,762 |
| 9 | poisson3Db | 85,623 | 2,374,949 |
| 10 | airfoil_2d | 14,214 | 259,688 |
| 11 | viscoplastic2 | 32,769 | 381,326 |
| 12 | xenon1 | 48,600 | 1,181,120 |
| 13 | xenon2 | 157,464 | 3,866,688 |
| 14 | wang3 | 26,064 | 177,168 |
| 15 | wang4 | 26,068 | 177,196 |
| 16 | ecl32 | 51,993 | 380,415 |
| 17 | sme3Da | 12,504 | 874,887 |
| 18 | sme3Db | 29,067 | 2,081,063 |
| 19 | sme3Dc | 42,930 | 3,148,656 |
| 20 | epb1 | 14,734 | 95,053 |
| 21 | epb2 | 25,228 | 175,027 |
| 22 | epb3 | 84,617 | 463,625 |

The symmetric matrices are used for OpenATI_EIGENSOLVE (Lanczos). The unsymmetric matrices are used for the other solvers.

The sparse matrix format for OpenATLib is CRS (compressed row storage). For the symmetric matrices, the lower part of the matrix is not stored in order to save memory space.

## 4.3 Experimental Condition of the Solvers

We evaluated two meta-solvers by using the following conditions.

- OpenATI_EIGENSOLVE
    - ➢ Standard eigenproblem
        - ✧ Restarted Lanczos method (Symmetric matrices).
        - ✧ Restarted Arnoldi method (Unsymmetric matrices).
    - ➢ End-user accuracy requirement: 1.0E-08.
    - ➢ The number of eigenvalues and eigenvectors: 10.
    - ➢ Maximum execution time: 1,000 seconds
- OpenATI_LINEARSOLVE
    - ● Linear equations
        - ✧ Restarted GMRES(m) method (Unsymmetric matrices).
        - ✧ BiCGStab method (Unsymmetric matrices).
    - ➢ End-user accuracy requirement: 1.0E-08.
    - ➢ Initial RHS vector of x: All elements are set to 1.
    - ➢ Initial guess: All elements are set to 0.
    - ➢ Preconditioner: ILU(0).
    - ➢ Maximum execution time: 1,000 seconds.
- AT for Restart Frequency
    - ➢ The difference from Figure 2 is summarized as follows.
        - ✧ Call OpenATI_DAFRT in every restarting.
        - ✧ If IRT= 1, then set MSIZE=MSIZE+5.

## 4.4 Results

### 4.4.1 SpMV Performance

#### 4.4.1.1 Symmetric Case

The typical performance of OpenATI_DSRMV is shown in Figure 5. According to the figure, by using the non-zero element based method with omission of the zero-element computation for vector reduction (S3), we can establish scalable speedup, whereas the conventional method (S1) cannot establish scalable speedup. In the matrix of gyro in 16 threads, the conventional method (S1) only attained 0.5 GFLOPS, but the proposed (S3) attained 7.0 GFLOPS. The speedup factor was 14.

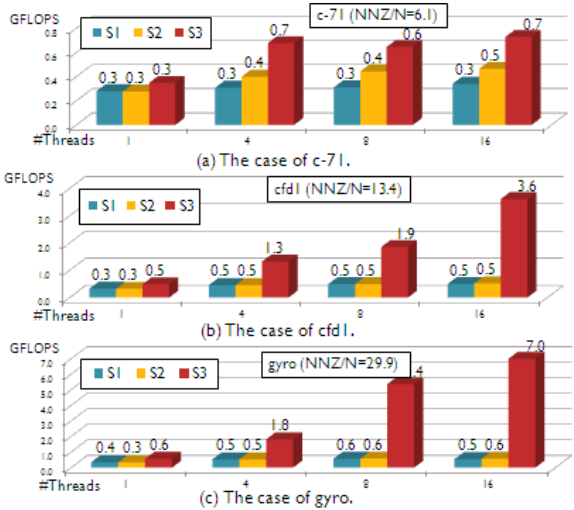

Figure 5. Typical Performance of OpenATI_DSRMV on the T2K (1 node).

#### 4.4.1.2 Unsymmetric Case – General

The typical performance in OpenATI_DURMV is shown in Figure 6. According to the figure, by using the non-zero element based method (U2), we can establish much speedup compared to that by the conventional method (U1). In the matrix of epb3 in 16 threads, the conventional (U1) only attained 3.8 GFLOPS, whereas the proposed (U2) attained 10.7 GFLOPS. The speedup factor was 2.8.

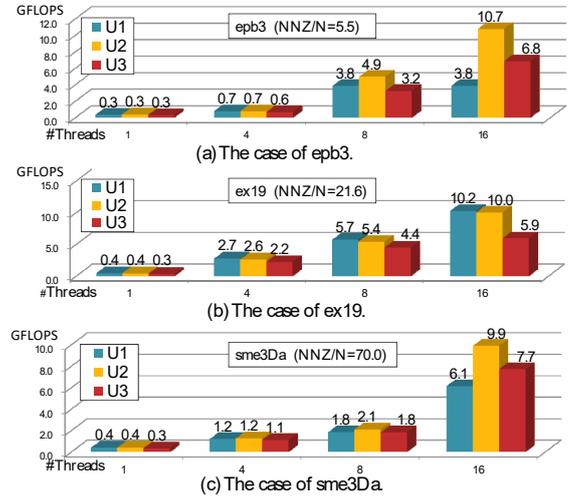

Figure 6. Typical Performance of OpenATI_DURMV on the T2K (1 node).

#### 4.4.1.3 Unsymmetric Case – Very Unbalanced

In the previous experiment, BSS (U3) offered no advantages. The matrix for which SS is superior to the conventional method has a very unbalanced number of non-zero elements per row. To show this performance, we picked a very unbalanced matrix from the UF collection. The matrix is IBM/EDA trans4, which has a dimension of 116,835, and 749,800 non-zero elements. The mean number of non-zero elements per row is 6.5, the maximum number is 114,910, and the minimum is 1. In addition, the SS approach has a performance parameter for the number of segment vectors. We denote the number of segments by JL. Figure 7 shows the performance in 16 threads with JL varying from 8 to 256.

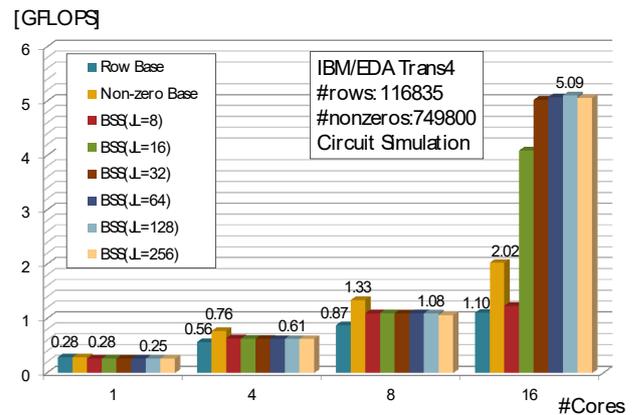

Figure 7. Performance of BSS (U3) in 16 threads of the T2K (1 node).

In Figure 7, the performance of BSS depends on the increase of the number of threads. In 16 threads, the BSS with JL=128 reached 5.09 GFLOPS, whereas the conventional row based method (U1) reached 1.10 GFLOPS. The speedup factor was 4.62.

### 4.4.2 Restart Frequency Adjustment

The adjustment of the restart frequency via OpenATI_DAFRT is one of the originalities for OpenATLib. In this section, a typical effect for the AT of the restart frequency is shown.

In OpenATLib, library developers determine the strategy for adjustment of the restart frequency. In the current implementation of Xabclib, we use the strategy shown in Figure 2 in this experiment.

Figure 8 shows the AT effect for OpenATI_EIGENSOLVE (Arnoldi) by using OpenATI_DAFRT. The X-axis shows the fixed restart frequency for all iterations except for the AT. In the AT, the restart frequency is dynamically changed according to the timings of the restart, found by monitoring the residual norms. In this experiment, the implementation of SpMV was fixed for the non-zero elements based method (U2), and the re-orthogonalization was fixed for the BCGS. The maximum number of iterations was set to 10,000.

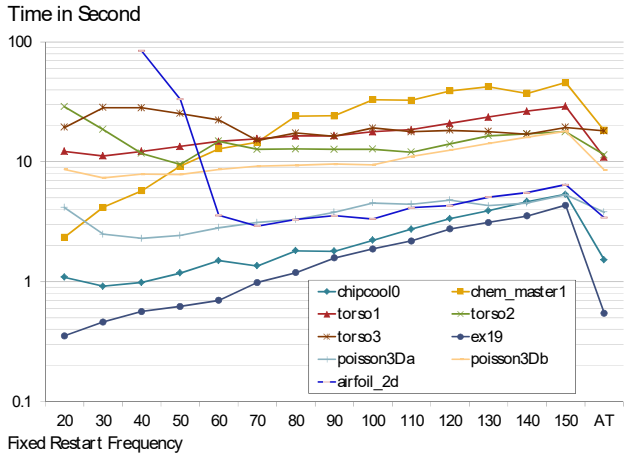

Time in Second

**Figure 8. Auto-tuning Effect of Restart Frequency for OpenATI_EIGENSOLVE(Arnoldi) on the T2K (16 threads).**

In Figure 8, the execution with the AT for torso1 was the fastest. In airfoil_2d, the fixed restarts of 20 and 30 did not converge. With the AT, the iteration converged for airfoil_2d. However, execution with the AT for chem_master_1 was 18.3 seconds, but the case of 20 restarts was 2.35 seconds. This is a 7.78-fold decrease in speed. This implies that the current restart strategy, which depends on the end-user description, is good for difficult problems. A difficult problem means that the problem requires 50 or more iterations/frequencies for the restarting frequency. Improvement of the strategy of the restart frequency for easy problems, which require less than 30 frequencies, is a topic of future work.

### 4.4.3 Effect of the Numerical Computation Policy

#### 4.4.3.1 Execution Time Policy

Figure 9 shows the effect of the AT for OpenATI_EIGENSOLVE (Arnoldi) from the viewpoint of execution time. The X-axis shows the execution time for each policy on matrices from the UF collection.

Figure 9 shows that almost all executions closed by using the time policy. But the cases of epb1 and epb2 indicate the effectiveness of the time policy. In the case of epb1, the execution with the time policy was 10× faster than that with the memory policy. This is because a large restarting frequency is required to establish fast convergence in epb1, but it requires much memory. This is totally against the memory policy. For this reason, the execution time with the memory policy was dramatically increased.

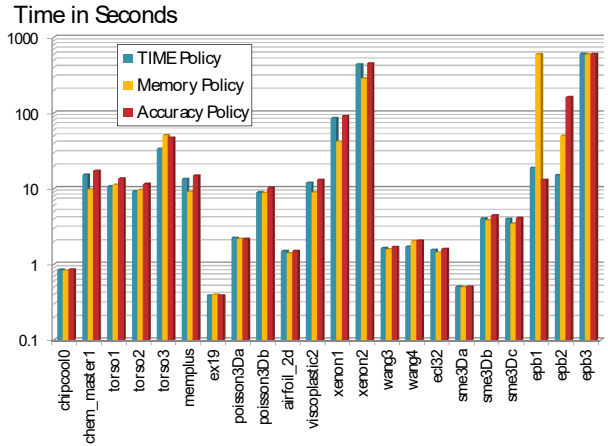

Time in Seconds

**Figure 9. Effect of the AT for OpenATI_EIGENSOLVE (Arnoldi) from the Viewpoint of Execution Time.**

#### 4.4.3.2 Memory Policy

Figure 10 shows the effect of the AT for OpenATI_EIGENSOLVE (Arnoldi) from the viewpoint of memory space.

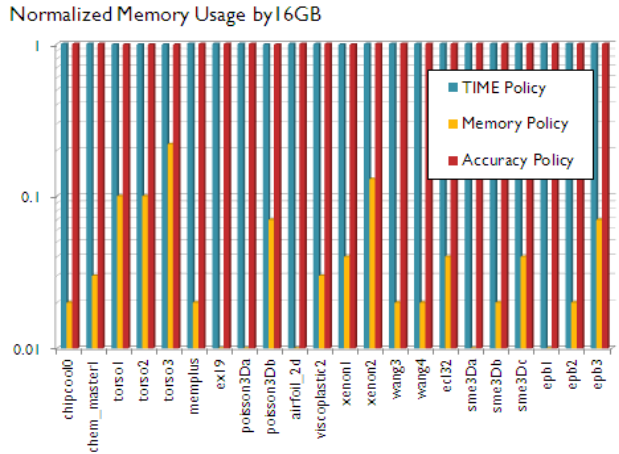

Normalized Memory Usage by16GB

**Figure 10. Effect of the AT for OpenATI_EIGENSOLVE (Arnoldi) from the Viewpoint of Memory Space. The value is normalized by 16 GB.**

In Figure 10, by using the memory policy, the memory space was dramatically reduced in comparison with the other two policies. The averaged reduction of memory space for all matrices reached 1/45 of the time and accuracy policies. This is because, to keep the maximum restart frequency, all remaining memory space, which had maximum 16 GB in this experiment, was consumed for the time and accuracy policies.

### 4.4.3.3 Accuracy Policy

Figure 11 shows the effect of the AT for OpenATI_EIGENSOLVE (Arnoldi) from the viewpoint of accuracy.

In the experiment of Figure 11, the end-user requirement was set to 1.0E-8. According to the figure, fault convergence occurred in epb2. In epb2, execution with the accuracy policy satisfied the accuracy requirement from the end-user. In summary, for the end-user accuracy requirement, 20 matrices were satisfied by the time policy, 19 matrices by the memory policy, and 21 matrices by the accuracy policy.

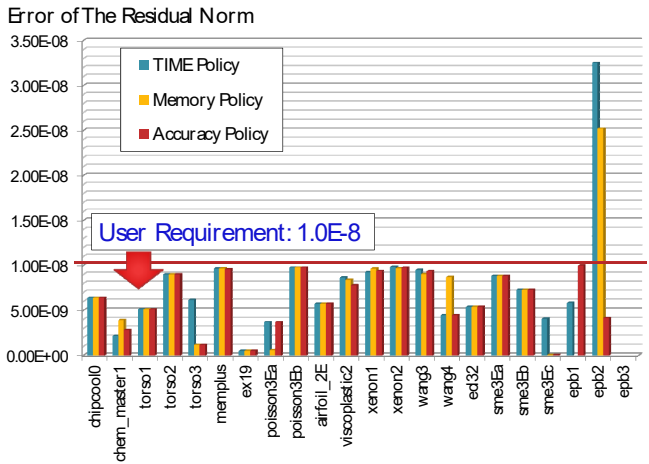

**Figure 11. Effect of the AT for OpenATI_EIGENSOLVE (Arnoldi) from the Viewpoint of Accuracy. The Y-axis Shows the Maximum 2-norm for Residual Vectors of the Standard Eigenproblem.**

**Effect of the Accuracy Policy for Other Solvers**

Table 4 shows the number of matrices that satisfied the accuracy of the end-user requirement for 22 kinds of test matrices from the UF collection (unsymmetric matrices).

**Table 4. Number of Matrices Satisfying the Accuracy of the End-user Requirement.**

| Solver Name | Execution Time Policy | Memory Amount Policy | Accuracy Policy |
|---|---|---|---|
| EIGENSOLVE (LANCZOS) | 20 | 20 | **21** |
| EIGENSOLVE (ARNOLDI) | 20 | 19 | **21** |
| LINEARSOLVE (GMRES) | **15** | 14 | **15** |
| LINEARSOLVE (BICGSTAB) | **14** | **14** | **14** |

Table 4 indicates that the accuracy policy established better accuracy from the viewpoint of the satisfied end-user requirement for all solvers compared to that of the other two policies. In other words, the AT accuracy policy can better satisfy the requirements compared to the conventional AT policy.

## 5. CONCLUSION

In this paper, we proposed general APIs, named OpenATLib, for auto-tuning (AT). OpenATLib is designed to establish the reusability of AT functions. This is the first originality of this paper.

The second originality is to develop high-performance sparse matrix-vector multiplication (SpMV). To establish this, we proposed the following new implementations: (1) non-zero based decomposition; (2) zero-elements computation omission for vector reduction; (3) branchless segmented scan (BSS). According to the performance evaluation, we obtained the following maximum speedups in comparison to those of conventional implementations: (1) 14× by the non-zero based method and by the zero-elements computation omission for symmetric SpMV; (2) 4.62× by BSS.

The third originality is to develop a numerical computation policy for sparse iterative solvers. Conventional AT focuses on execution time only. We established a strategy to optimize memory space and accuracy as numerical computation policies. By using the new numerical computation policies, we obtained (1) an averaged 1/45 memory space reduction; (2) avoidance of the "fault convergence" situation caused by conventional iterative solvers.

The fourth originality is implementation of the general AT strategy for a restart frequency adjustment by using the MM ratio. The strategy is independent of numerical algorithms, since the strategy only monitors crucial values in the solver at run-time. In addition, the strategy does not rely on linear algebra theory. By restarting the AT, we can stabilize the solver, even if we need to solve "difficult" problems, which require a high restart frequency, such as 50 iterations or more.

To establish the ultimate auto-tuned iterative library, we need to develop many additional AT functions, such as the selection between BiCGStab and GMRES, and the selection of orthogonalization algorithms, including MGS and DGKS. The preconditioner selection and auto-tuning of performance parameters are also crucial developments. This includes the selection of ILU or Block Jacobi, the selection of ILUT and its fill-in depths. For SpMV implementation, an AT for dynamic data transformation is an important function for obtaining much higher performance, including, for example, dynamic data transformation and corresponding implementations of SpMV, such as CRS to ELL or COO formats. The blocking factors of SpMV, such as the adjustment of block length for the BCSR format, are important. The above topics are considered for future work of the Xabclib project.

## 6. ACKNOWLEDGMENTS

This study is supported by "Seamless and Highly Productive Parallel Programming Environment for High-Performance Computing," granted by the Ministry of Education, Culture, Sports, Science and Technology (MEXT), Japan.